\documentclass[prl,twocolumn,superscriptaddress,amsfonts,amssymb,showpacs,a4]{revtex4}
\usepackage{graphicx}
\usepackage{dcolumn}
\usepackage{times}
\usepackage{amsmath}

\usepackage{color}
\renewcommand{\vec}[1]{\textbf{\textit{#1}}}

\begin{document}
\title{Interplay of Dirac fermions and heavy quasiparticles in solids}

\author{M. H\"oppner}
\affiliation{Max Planck Institute for Solid State Research, Heisenbergstrasse 1, D-70569 Stuttgart, Germany}
\affiliation{Institute of Solid State Physics, Dresden University of Technology, D-01062 Dresden, Germany}

\author{S. Seiro}
\affiliation{Max Planck Institute for Chemical Physics of Solids, D-01187 Dresden, Germany}

\author{A. Chikina}
\affiliation{Institute of Solid State Physics, Dresden University of Technology, D-01062 Dresden, Germany}

\author{A. Fedorov}
\affiliation{Leibniz-Institut f\"ur Festk\"orper- und Werkstoffforschung Dresden, P.O. Box 270116, D-01171 Dresden, Germany}

\author{M. G\"uttler}
\affiliation{Institute of Solid State Physics, Dresden University of Technology, D-01062 Dresden, Germany}

\author{S. Danzenb\"acher}
\affiliation{Institute of Solid State Physics, Dresden University of Technology, D-01062 Dresden, Germany} 

\author{A. Generalov}
\affiliation{Institute of Solid State Physics, Dresden University of Technology, D-01062 Dresden, Germany}

\author{K. Kummer}
\affiliation{European Synchrotron Radiation Facility, 6 Rue Jules Horowitz, Bo\^{\i}te Postale 220, F-38043 Grenoble Cedex, France}

\author{S.L. Molodtsov}
\affiliation{European XFEL GmbH, Albert-Einstein-Ring 19, D-22671 Hamburg, Germany}

\author{Yu. Kucherenko}
\affiliation{Institute of Solid State Physics, Dresden University of Technology, D-01062 Dresden, Germany}
\affiliation{Institute for Metal Physics, National Academy of Sciences of Ukraine, UA-03142 Kiev, Ukraine}

\author{C. Geibel}
\affiliation{Max Planck Institute for Chemical Physics of Solids, D-01187 Dresden, Germany}

\author{V. Strocov}
\affiliation{Swiss Light Source, Paul Scherrer Institute, CH-5232 Villigen-PSI, Switzerland}

\author{M. Shi}
\affiliation{Swiss Light Source, Paul Scherrer Institute, CH-5232 Villigen-PSI, Switzerland}

\author{M. Radovic}
\affiliation{Swiss Light Source, Paul Scherrer Institute, CH-5232 Villigen-PSI, Switzerland}
\affiliation{Laboratory for Synchrotron and Neutron Spectroscopy-ICMP, Ecole Polytechnique Federale de Lausanne, CH-1015 Lausanne, Switzerland}

\author{T. Schmitt}
\affiliation{Swiss Light Source, Paul Scherrer Institute, CH-5232 Villigen-PSI, Switzerland}

\author{C. Laubschat}
\affiliation{Institute of Solid State Physics, Dresden University of Technology, D-01062 Dresden, Germany}

\author{D. V. Vyalikh}
\affiliation{Institute of Solid State Physics, Dresden University of Technology, D-01062 Dresden, Germany}

\date{\today}
\begin{abstract}

Many-body interactions in crystalline solids can be conveniently described in terms of quasiparticles with strongly renormalized masses as compared to those of non-interacting particles. Examples of extreme mass renormalization are on the one hand graphene, where the charge carriers obey the linear dispersion relation of massless Dirac fermions, and on the other hand heavy-fermion materials where the effective electron mass approaches the mass of a proton. Here we show that both extremes, Dirac fermions like they are found in graphene and extremely-heavy quasiparticles characteristic for Kondo materials, may not only coexist in a solid but can undergo strong mutual interactions. Using the example of EuRh$_2$Si$_2$ we explicitly demonstrate that these interactions can take place at the surface and in the bulk. The presence of the linear dispersion is imposed solely by the crystal symmetry while the existence of heavy quasiparticles is caused by the localized nature of the 4f states. 

\end{abstract}


\maketitle
The idea that collective excitations, e.g. lattice vibrations, and many-body phenomena, e.g. electron correlation, can be described in terms of a rescaled single-particle problem~\cite{1} was a major breakthrough in condensed matter physics. The solutions of such models are quasiparticles which relate the many-body interactions to the properties of a virtual single particle. This concept can provide an understanding of complex phenomena like e. g. the exotic low-temperature physics of heavy-fermion systems–intermetallics synthesized on the basis of $4f$ or $5f$ elements – where the conduction electrons act as particles with huge masses, comparable to that of a proton~\cite{2,3}. The large effective masses are reflected in the low temperature properties of the material, for instance, by large values of the specific heat or an increase of electric resistivity below the so-called Kondo temperature. In angle-resolved photoemission (ARPES) data, the respective states can be identified as flat running states with an almost complete absence of dispersion. A diametrically-opposed example are charge carriers in graphene, that behave as massless Dirac fermions~\cite{4,5,6,7,8}, which are, like photons, characterized by a linear dispersion relation~\cite{6}. The results of A.~Geim and K.~Novoselov explicitly demonstrate a correlation of this linear dispersion with other unique properties of graphene like the large opacity, the huge charge carrier mobility, and the anomalous quantum-Hall effect~\cite{4,5,7,8}. Here we report that both extremes, massless and ultra-heavy fermionic quasiparticles may not only coexist in a solid but even exhibit mutual interaction and form composite quasiparticles. 

To this end, we studied the electronic structure of the layered material EuRh$_2$Si$_2$ by means of ARPES and density functional theory (DFT) calculations. EuRh$_2$Si$_2$ is a stable-valent representative of the broad family of famous heavy-fermion compounds like YbRh$_2$Si$_2$, CeCu$_2$Si$_2$ and URu$_2$Si$_2$~\cite{9,10}. Around the \mbox{$\overline{\Gamma}$-point} of the Brillouin zone (BZ) we observe a sharp, linearly dispersing band of Rh 4$d$-derived states. Its conical shape is directed towards the Fermi level ($E_\mathrm{F}$) and it passes through an Eu 4$f^6$ ($^7$F$_\mathrm{J}$) excited state multiplet with whom it interacts. A similar interaction is maintained when going from the surface into the bulk as proven by using soft X-ray ARPES with high bulk sensitivity. The discovery of quasiparticles simultaneously formed by massless and heavy-fermion states is quite remarkable because species showing either Dirac or heavy fermions typically exhibit opposing material properties.

The linear band dispersion around $\overline{\Gamma}$ seems to exist in various members of the tetragonal body-centered ThCr$_2$Si$_2$ (122) family. In the isostructural Kondo lattice compound YbRh$_2$Si$_2$, a respective band has been previously found but there the apex of the Dirac-like cone is slightly deformed by hybridization with 4$f$~states at the Fermi level~\cite{10,11,12}. A similar interplay between heavy 4$f$ and massless 4$d$ quasiparticles also occurs in EuRh$_2$Si$_2$ leading to a pronounced non-crossing behaviour that is explicitly reminiscent to that of polariton generation in optics where a light cone interacts with an almost non-dispersive optical phonon branch~\cite{13}. In contrast to YbRh$_2$Si$_2$ the linear band dispersion is preserved at $E_\mathrm{F}$ in EuRh$_2$Si$_2$ due to the absence of energetically low-lying 4$f$~excitations. Because of its hybridization with the 4$f$~multiplet the linearly dispersing band accumulates a certain weight of 4$f$~character which it carries all the way up to the Fermi level region. A computational analysis provides a hint to spin polarization of the linear band. This is in close analogy to what is observed for topological insulators~\cite{14,15,16} where respective linear dispersive surface states bridge the band gap and lead to metallic behaviour in the surface layer. However, the linear state in EuRh$_2$Si$_2$ is probably not protected by time-reversal symmetry like in topological insulators since EuRh$_2$Si$_2$ is a metal and the state is also found in the bulk. Thus it may be closer to the case of graphene where the linear dispersion emerges due to the lattice symmetry.

\begin{figure}
  \centering
  \includegraphics[width=0.46\textwidth]{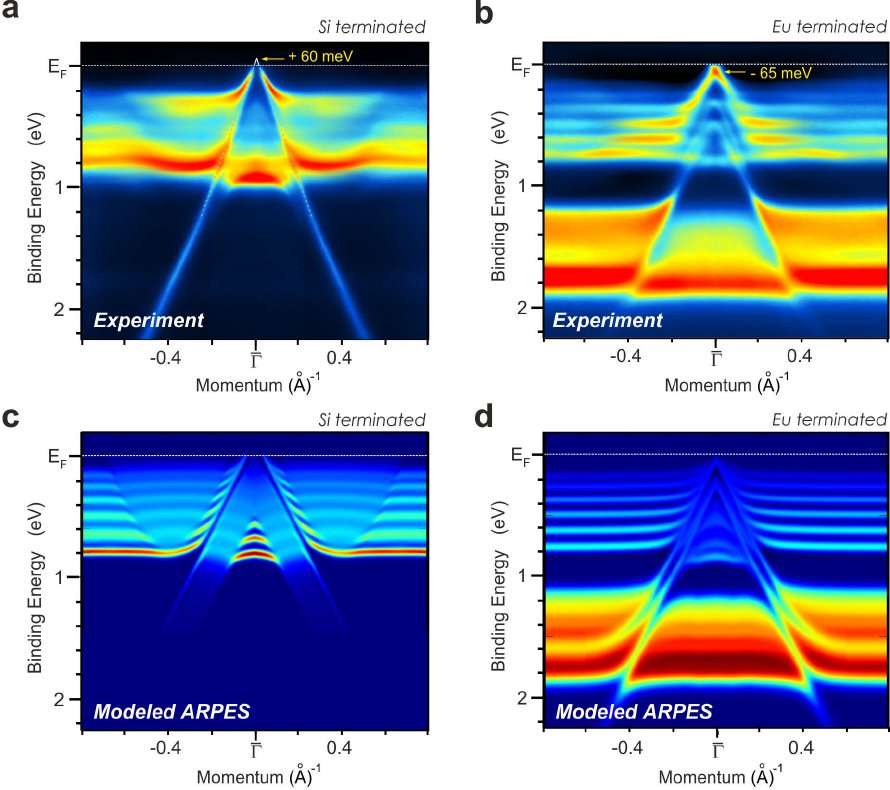}
  \caption{Experimental ARPES spectra taken along the $\overline{\Gamma}$--$\overline{\mathrm{X}}$-direction with the photon energy 120~eV, i.e.~\textit{sensitive} to the Eu~4$f$ emission while the Rh~4$d$ states are strongly suppressed by the Cooper minimum, for \textbf{(a)} Si~terminated and \textbf{(b)} Eu~terminated surfaces. The simulated model ARPES spectra for the same surfaces are presented in \textbf{(c)} and \textbf{(d)}, respectively.}
  \label{fig1}
\end{figure}

EuRh$_2$Si$_2$ usually cleaves along the Eu~layers leaving behind surfaces terminated either by Si or Eu~atoms. In figs.~\ref{fig1}a and b we compare the electron band structure as seen by ARPES along the $\overline{\Gamma}$--$\overline{\mathrm{X}}$-direction for the two possible surface terminations after cleaving. The band topologies look rather different in both cases: The 4$f$~electron emission from bulk Eu~atoms, visible as the narrow, string-like 4$f^6$~final state multiplet between 0.2 and 0.8~eV binding energies (BE), is detected for both surfaces, while the presence of Eu~atoms at the surface gives rise to an additional broad 4$f$~signal between 1.2 to 2~eV~BE. 

The strong localization of 4$f$~electrons leads to the observed almost dispersion-less behaviour corresponding to the limit of free ions or -- rephrased in the band structure picture -- infinitely heavy quasiparticles. These "horizontal bands” are disturbed at those points in \vec{k}-space, where they are crossed by the linearly dispersing valence band. Where "heavy" and "light" electron bands would cross, the energy degeneracy is lifted by hybridization leading to the formation of small energy gaps and a mixing of the Eu~4$f$ and Rh~4$d$ states that is explicitly reflected in the ARPES data. The linear bands clearly hybridize with all J-terms of the Eu~4$f$ multiplet. The resulting composite states exhibit a non-vanishing $f$~character and are, thus, nicely seen at the chosen photon energy of 120~eV where 4$f$~emissions dominate. Effectively massless quasiparticles with finite 4$f$~character appear in this way even at the Fermi energy and form a strong contrast to the heavy 4$f$-derived quasiparticles usually observed in Kondo systems.

Looking closely at the $f$-$d$~hybrid band, running towards $E_\mathrm{F}$ in a quasi-conical shape, we can make important observations. The energy position of the cone apex depends on the surface termination: for Si~termination it appears $\approx 60$~meV above the Fermi level while for Eu~termination it is located $\approx 65$~meV below. Upon changing the photon energy no notable dispersion in $k_\mathrm{z}$-direction is observed. These properties are well reproduced in band structure calculations. To this end we used a slab geometry where 55~(57)~atomic layers parallel to the surface for Si~(Eu)~termination are considered. That allows to discriminate between surface and bulk effects by the localization of charge density. Treating the 4$f$~electrons as core states the linearly dispersing bands are reproduced by electron states of almost two-dimensional character which are located at the surface.

\begin{figure}
  \centering
  \includegraphics[width=0.46\textwidth]{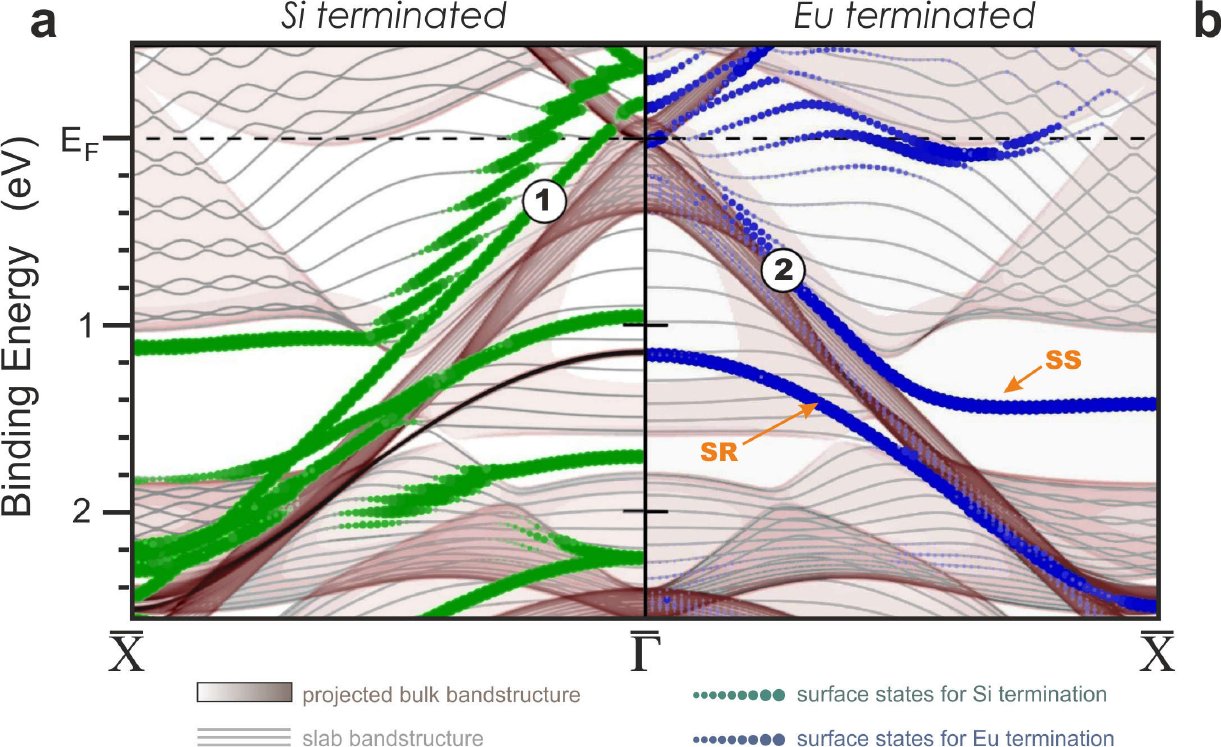}
  \caption{LDA~based band structure calculations for a surface covered by \textbf{(a)} Si and \textbf{(b)} Eu atoms. The maroon marked area corresponds to the surface-projected bulk band structure whereas the grey lines are the result of a calculation for a slab model. The green (blue) dots scale with ratio of localization: if the charge density is mainly located at the surface, then the size of the respective dot representing the eigenfunction is at its maximum whereas for equally distributed charge-density it vanishes.}
  \label{fig2}
\end{figure}

Understanding the interaction between localized 4$f$ and itinerant electrons requires information on the strength of the hybridization which is intimately related to the symmetry of coupling orbitals. Within a simple hybridization model one may assume that the hybridization strength is proportional to the $f$-character of the unhybridized valence states at the site of the Eu~atoms. 

Respective information may be obtained from our slab calculations. It can be used then to simulate the ARPES spectra for both surface terminations of the EuRh$_2$Si$_2$ crystal. The results are shown in figs.~\ref{fig1}c and \ref{fig1}d which qualitatively reproduce the key features of our experimental ARPES data. The linear dispersion of the Rh bands obtained in the slab calculations are mainly preserved in the hybridization model, modified by hybridization gaps and the crossover to the Eu 4$f^6$~final state multiplet. Additionally, the hybridization model yields the transfer of Eu~4$f$ spectral weight to the conically shaped valence band confirming the conclusions drawn from the experimental data.

In fig.~\ref{fig2} we show the unhybridized bulk and surface bands for both surface terminations as computed from the slab model. In the calculations the localized 4$f$~states were treated as core states. Green and blue dotted lines represent surface bands, which are shifted in energy with respect to the bulk bands, and surface states, which are located inside a bulk band gap. The thicknesses of these lines scale with the surface to bulk ratio. The bulk band-structure projected into the surface BZ~\cite{17} is plotted in shades of maroon. There is remarkable agreement between the experimentally derived electron bands and the computed band dispersions for both terminations. In particular, the surface bands with linear dispersion at the $\overline{\Gamma}$-point are well reproduced, including the energy difference of the cone apex position between the Eu and Si~termination. In addition, there is a finite gap of 200~meV between the upper and the lower cone for both surface configurations.

Group-theory considerations on two-dimensional lattices show that the quasiparticle dispersion around high-symmetry points is mainly determined by the symmetry of the space group of the crystal lattice~\cite{18,19}. Similar predictions for the space groups in three dimensions have not been carried out yet, but the arrangement of Rh~atoms at the surface in the layered structure of EuRh$_2$Si$_2$ with tetragonal symmetry can up to some extent be approximated by a squared lattice for which in fact the appearance of linear electron dispersion was predicted. 

\begin{figure}
  \centering
  \includegraphics[width=0.46\textwidth]{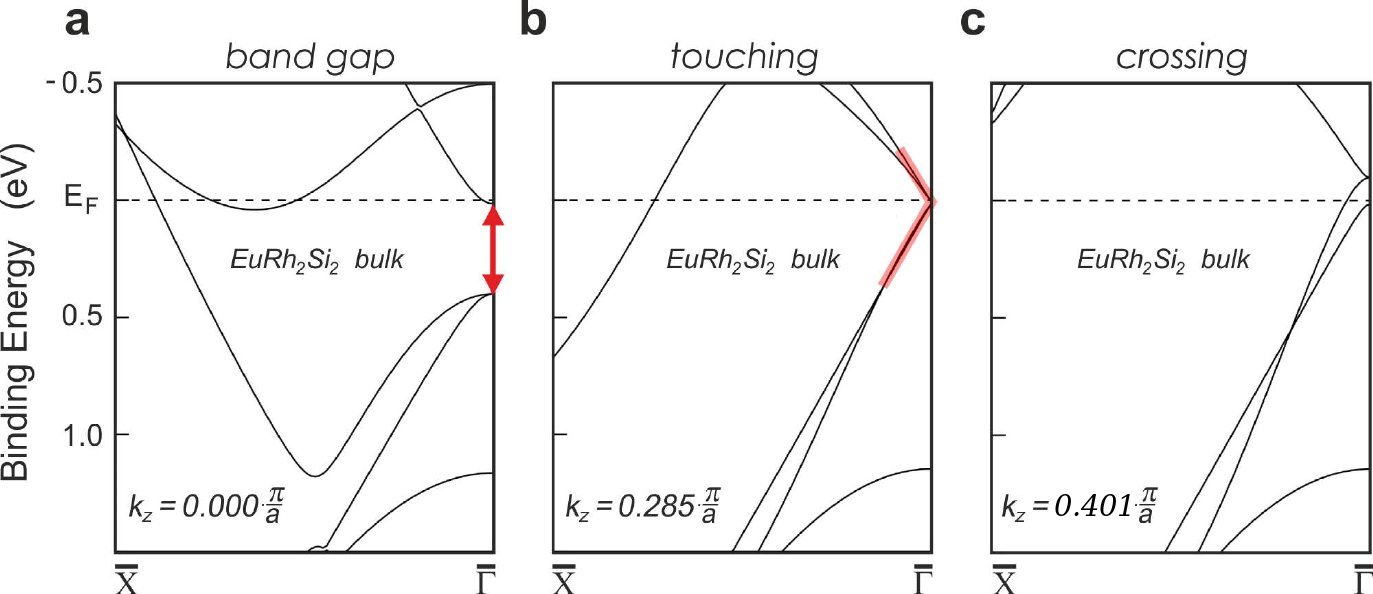}
  \caption{LDA~derived bulk band structure parallel to the $\overline{\Gamma}$--$\overline{\mathrm{X}}$-direction for several planes $k_\mathrm{x} \times k_\mathrm{y}$ defined by $k_\mathrm{z}$. There are few different bands in the region around the $\overline{\Gamma}$-point close to $E_\mathrm{F}$. Going from the middle of the BZ to the top (bottom), one notices that they form a quasi-linear dispersion (2D) at $k_\mathrm{z} = \pm 0.285 \cdot \pi/a$, which gets enhanced as well as shifted in energy at the surface.}
  \label{fig3}
\end{figure}

The layered character of the compound poses the question whether similar phenomena as observed in the surface region may also be expected for the bulk. The projected bulk band structure shows indeed a band that coincides roughly with the linear surface band and reveals a weakly three-dimensional (3D) character (dark maroon color in fig.~\ref{fig2}). This becomes more evident in fig.~\ref{fig3}, where we show cuts at several $k_\mathrm{z}$ through the bulk BZ, parallel to the $\overline{\Gamma}$--$\overline{\mathrm{X}}$-direction. For $k_\mathrm{z} = 0.285~\pi$/a the bands in the vicinity of the $\overline{\Gamma}$-point become degenerate and reveal linear dispersion while for other values of $k_\mathrm{z}$ the band dispersions deviate slightly from this linear slope. Thus, similar hybridization phenomena as observed in the surface region may in fact be found in the bulk, too. To prove this, we performed an ARPES experiment in the soft X-ray photon energy range where the photoelectron escape depth is increased and the extreme surface sensitivity of XUV ARPES is partly overcome. 

\begin{figure}
  \centering
  \includegraphics[width=0.46\textwidth]{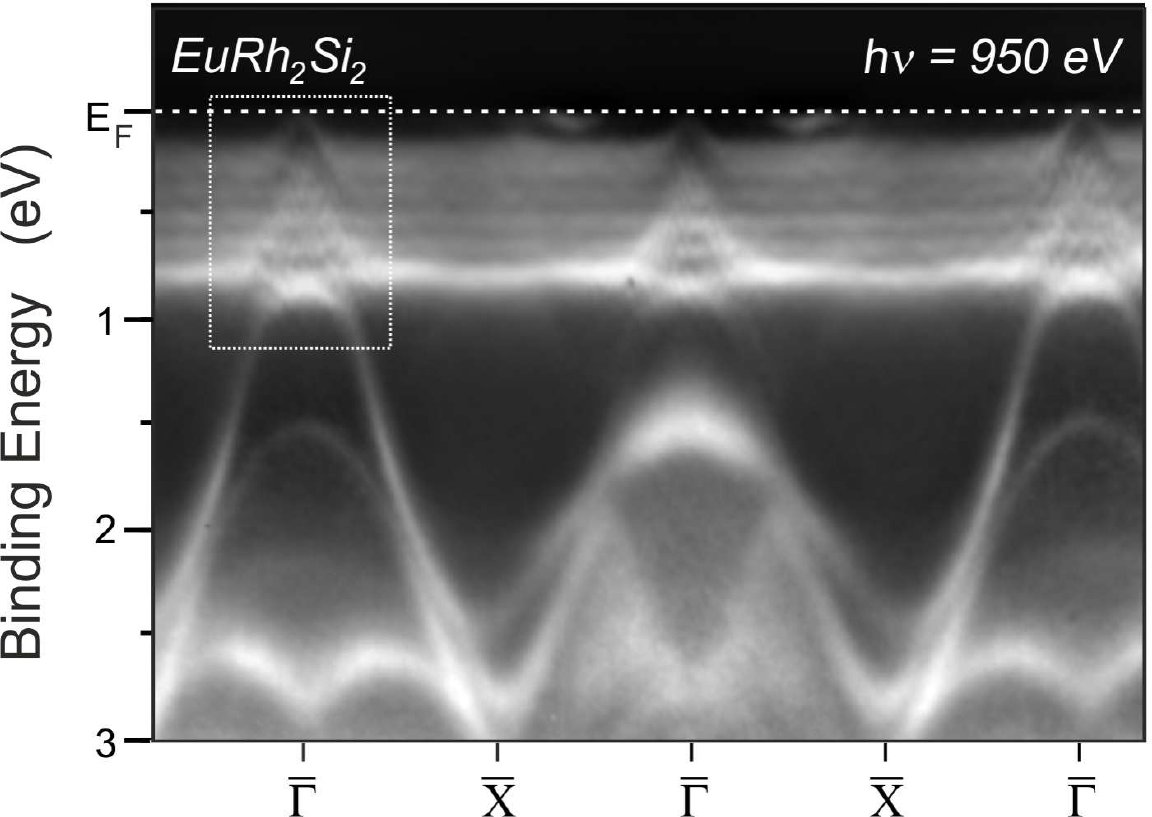}
  \caption{Experimental ARPES data of EuRh$_2$Si$_2$ recorded using 950~eV photons, which allows to suppress the emission from the surface states and thus to disentangle the bulk electron structure of the system. The dotted rectangular shows the region where conically shaped Rh bands with their apex close to the Fermi level are coupled with the bunch of ``horizontal'' bands from the bulk Eu~4$f^6$ multiplet. }
  \label{fig4}
\end{figure}

In fig.~\ref{fig4} we show an ARPES map of EuRh$_2$Si$_2$ taken at $\mathrm{h}\nu=950$~eV that extends over the $\overline{\Gamma}$-points of three adjacent Brillouin zones. As we know from respective results for the isostructural compound YbRh$_2$Si$_2$~\cite{20} the contribution of the first subsurface Eu~layer amounts to about only 20\% of the total 4$f$~emission at this photon energy range while it amounts to about 60\% in the data shown in fig.~\ref{fig1}. In fact, scanning the beam across the whole surface of freshly cleaved EuRh$_2$Si$_2$ we did not detect a sign of surface-related electron states. In particular, we never found the surface Eu~4$f$~feature at $\approx 1.6$~eV~BE although the atomic cross section for Rh~4$d$ and Eu~4$f$ is of the same magnitude at 950~eV photon energy. In the XUV photon energy range it was always seen for some regions of the sample surface. The absence of surface related features confirms the bulk sensitivity in the soft \mbox{X-ray} range. In the region around the $\overline{\Gamma}$-points we observe well-defined bands that approach $E_\mathrm{F}$ in a quasi-linear way similar to the surface band discussed above. Furthermore, we also detect the string-like Eu~4$f^6$ final-state multiplet which again hybridizes with the linearly dispersive bands. The strong similarities between the surface-sensitive XUV and the more bulk sensitive soft X-ray data is owed to the layered structured of the material and the weak three-dimensional character of the underlying Rh 4$d$ band. Although the energy and momentum resolution in the soft X-ray range are not as high as in the XUV they still allow the conclusion that ultra-light and heavy quasiparticles coexist for particular cuts through the 3D~Brillouin zone of the material, too, and are subject to the same hybridization phenomena as their surface counterparts. 

In summary, we have demonstrated that massless and heavy fermion quasiparticles may not only coexist but can also strongly interact in solids. Applying XUV and soft \mbox{X-ray} ARPES on EuRh$_2$Si$_2$ single crystals we observed linearly dispersing Rh~4$d$-derived bands, corresponding to massless Dirac quasiparticles. Their conical shape is mainly imposed by the in-plane square symmetry of the layered compound. These linear bands are interfered by flat, dispersionless Eu~4$f$ states corresponding to quasiparticles with extremely large effective masses. At the regions of intersection of both kinds of bands hybridization gaps as well as strong admixture of the 4$f$~electron states to the Dirac fermions are detected. This unambiguously demonstrates a strong interplay between these two opposing limits of mass-renormalized fermions. Similar observations as for the surface were made for the bulk of the crystal. Since the symmetry of the layered (122) compounds dictates the formation of ultra-light quasiparticles, one may anticipate that details of their interactions with 4$f$~states could play a role in the emergence of unusual thermodynamic properties like Kondo~physics or quantum criticality in other compounds of this family.

\end{document}